\documentclass[sigconf,nonacm]{acmart}

\AtBeginDocument{%
  }

\setcopyright{acmlicensed}
\copyrightyear{2025}
\acmYear{2025}
\setcopyright{cc}
\setcctype{by-nc-nd}
\acmConference[UIST '25]{The 38th Annual ACM Symposium on User Interface Software and Technology}{September 28-October 1, 2025}{Busan, Republic of Korea}
\acmBooktitle{The 38th Annual ACM Symposium on User Interface Software and Technology (UIST '25), September 28-October 1, 2025, Busan, Republic of Korea}\acmDOI{10.1145/3746059.3747744}
\acmISBN{979-8-4007-2037-6/2025/09}

\usepackage{acmart-taps}
\usepackage{graphicx}
\usepackage{booktabs}
\usepackage{stfloats} 
\usepackage{amsthm}
\usepackage{svg}
\usepackage{svg-extract}
\usepackage{balance}

\usepackage{framed}
\usepackage{xcolor}

\newenvironment{tightframe}[1][black]{%
  \setlength{\FrameSep}{3pt}
  \setlength{\topsep}{2pt}
  \setlength{\fboxsep}{\FrameSep}%
  \setlength{\fboxrule}{2pt}
  \def\FrameCommand{\fcolorbox{#1}{white}}%
  \begin{framed}\begingroup\large
}{%
  \par\endgroup\end{framed}%
}

\begin{document}

\title{GraspR: A Computational Model of Spatial User Preferences for Adaptive Grasp UI Design}

\author{Arthur Caetano}
\email{caetano@ucsb.edu}
\orcid{0000-0003-0207-5471}
\affiliation{%
  \institution{University of California}
  \city{Santa Barbara}
  \state{CA}
  \country{USA}
}

\author{Yunhao Luo}
\email{yunhaoluo@ucsb.edu}
\orcid{0009-0004-6219-8021}
\affiliation{%
  \institution{University of California}
  \city{Santa Barbara}
  \state{CA}
  \country{USA}
}

\author{Adwait Sharma}
\email{as5339@bath.ac.uk}
\orcid{0000-0001-5676-3136}
\affiliation{%
  \institution{University of Bath}
  \city{Bath}
  \country{United Kingdom}
}

\author{Misha Sra}
\email{sra@ucsb.edu}
\orcid{0000-0001-8154-8518}
\affiliation{%
  \institution{University of California}
  \city{Santa Barbara}
  \state{CA}
  \country{USA}
}

\renewcommand{\shortauthors}{Caetano et al.}

\begin{abstract}
Grasp User Interfaces (grasp UIs) enable dual-tasking in XR by allowing interaction with digital content while holding physical objects. However, current grasp UI design practices face a fundamental challenge: existing approaches either capture user preferences through labor-intensive elicitation studies that are difficult to scale, or rely on biomechanical models that overlook subjective factors. We introduce GraspR, the first computational model that predicts user preferences for single-finger microgestures in grasp UIs. Our data-driven approach combines the scalability of computational methods with human preference modeling, trained on 1,520 preferences collected via a two-alternative forced choice paradigm across eight participants and four frequently used grasp variations. We demonstrate GraspR's effectiveness through a working prototype that dynamically adjusts interface layouts across four everyday tasks. We release both the dataset and code to support future research in adaptive grasp UIs.
\end{abstract}


\begin{teaserfigure}
  \includegraphics[width=\textwidth]{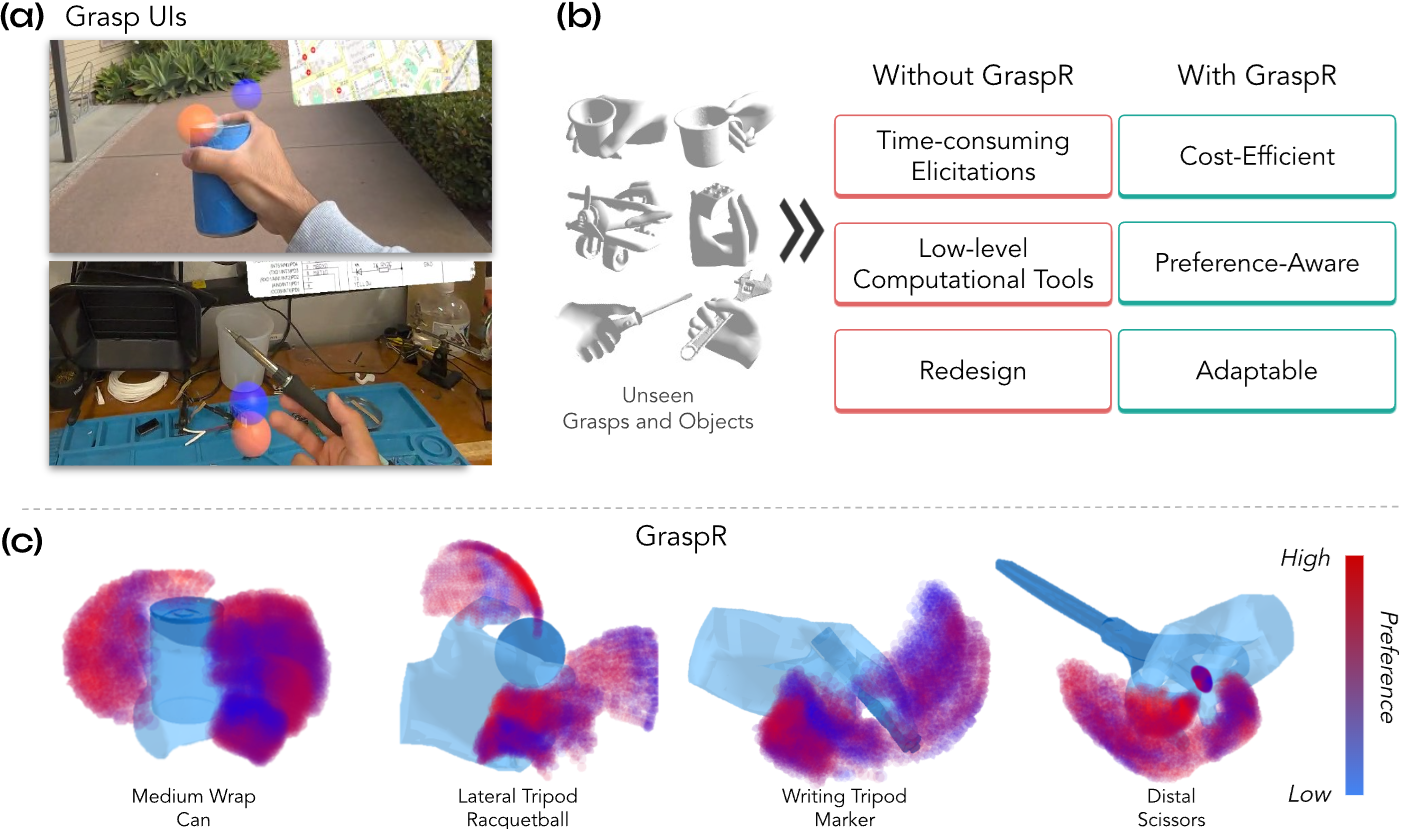}
  \caption{We present GraspR, a model to enable preference-aware adaptable grasp interfaces: (a) Grasp interfaces enable dual-tasking between virtual controls and affordances of physical objects. (b) Current grasp UI design methods rely on time-consuming elicitation studies, computational tools that lack alignment with high-level human factors, and require redesign to accommodate unseen grasp-object combinations. (c) GraspR is a model of user preference in single-finger grasp interaction that offers a cost-efficient approach for adaptive grasp UIs grounded on user preference data.}
  \Description{}
  \label{fig:teaser}
\end{teaserfigure}


\maketitle

\section{Introduction}

Grasp-based user interfaces (grasp UIs)~\cite{wolf2011taxonomy, sharma2019grasping, aponte2024grav, sharma2024graspui} enable users to seamlessly interact with virtual content while maintaining their grip on physical objects. These interfaces are particularly valuable in extended reality (XR), where users frequently engage in dual-task scenarios~\cite{ashbrook2010micro}, such as navigating virtual instructions with thumb-sliding gestures while holding tools~\cite{zhou2020gripmarks} or controlling media playback through an index-finger squeeze gesture while drinking from a mug~\cite{saponas2009enabling}. By seamlessly integrating physical and digital affordances, grasp UIs offer significant potential for improving productivity and enhancing usability in XR environments.

Despite their potential, designing adaptive grasp UIs that effectively conform to different grasp types and object geometries remains a significant challenge~\cite{bullock2013grasp, feix2015grasp, zuccotti2015everything}. Each object imposes unique constraints on finger mobility, UI visibility, and user comfort~\cite{wolf2011taxonomy, sharma2019grasping, aponte2024grav}. For instance, an interface optimized for sliding gestures on a pen may become cumbersome when users switch to a mug, due to differences in grasp types, finger movement range, and available surface for interaction. Similarly, an interface optimized for a coffee mug may stop working when the user switches to scissors. Or, factors such as surface utilization may produce layouts that conflict with other factors such as user preference~\cite{he2024adaptui, morris2010understanding}. The need to balance multiple, often conflicting design factors further amplifies the challenge of adaptive grasp UI design. Without such adaptivity, grasp UIs risk becoming rigid and impractical for real-world use.

Current grasp UI design approaches rely heavily on user elicitation studies~\cite{wobbrock2005maximizing, sharma2019grasping} or expert-driven iterative design~\cite{feix2015grasp, sharma2021solofinger, joshi2023transferable, sharma2024graspui}. These methods, while foundational, demand significant researcher investment, often requiring weeks of participant recruitment, multiple testing sessions per design variant, and extensive post-study analysis before yielding actionable insights. The resulting interfaces remain constrained by the experimental conditions that produced them, offering limited adaptability to unanticipated grasp variations or new object geometries that inevitably emerge in real-world use. This cost-intensive design workflow creates an inherent trade-off between comprehensiveness and practicality. Covering a wide range of grasp variations requires exponentially more study time, while practical design timelines demand faster iteration.

Computational design tools address some of the challenges of elicitation and expert-driven methodologies by reducing the time and labor required for iteration, enabling rapid exploration of design variants and adaptation to real-world variability. Systems such as AdapTutAR~\cite{huang2021adaptutar}, SituationAdapt~\cite{li2024situationadapt}, and context-aware MR interfaces~\cite{lindlbauer2019context} automate UI adaptation based on sensed context or user state, reducing reliance on static, manually-designed UIs. However, because these systems depend on complex sensing, reasoning, and optimization pipelines, they may struggle with noisy or incomplete real-world data. Their adaptability is limited by predefined models, rules, or heuristics, which restrict generalizability and responsiveness to unanticipated contexts. They can adjust in real-time, but have limited mechanisms to account for individual variation in comfort or preference. Furthermore, existing computational tools often emphasize biomechanical or task performance metrics, neglecting subjective criteria critical to creating natural and intuitive user experiences~\cite{wobbrock2009user, gajos2005preference, nielsen1994measuring}.

Recent simulation-based approaches offer alternatives to traditional adaptive UI design. For example, GraV~\cite{aponte2024grav} uses forward kinematics to evaluate reachability and joint movement, while Zhao et al.~\cite{zhao2024grip} assess grasp stability and accessible surface area. These systems enable objective evaluations of grasp feasibility, including displacement cost and trade-offs between stability and interactivity. While they provide valuable physical metrics, they do not explicitly consider user preferences, an aspect that may further enhance the intuitiveness and usability of adaptive grasp UIs.

To address the high time and labor costs of elicitation-based grasp UI design and the absence of user preference modeling in existing computational tools, we present GraspR (\textbf{Grasp} P\textbf{R}eferences), the first computational model of user preference for grasp UIs that generalizes to unseen grasp-object scenarios. GraspR is a pairwise classifier that predicts user preferences between alternative grasp UI interactions generated through simulation. By combining these predictions, we create a spatial preference map around the grasping hand, which serves as a basis for designing adaptive UIs that highlight preferred zones for more intuitive and comfortable interactions. Figure~\ref{fig:teaser}c shows a visualization of spatial preference predicted with GraspR. Our approach combines the scalability of simulation-based design with a predictive understanding of user preferences to bridge a key gap in the computational design of adaptive grasp UIs. Figure~\ref{fig:teaser}b visually summarizes how GraspR addresses limitations of grasp UI design methods. The following research question guides our work:

\aptLtoX{\begin{tightframenew}
How can we enable preference-aware adaptive grasp UIs that scale for a wide range of grasp-object interactions?
\end{tightframenew}}{
\begin{tightframe}[orange]
How can we enable preference-aware adaptive grasp UIs that scale for a wide range of grasp-object interactions?
\end{tightframe}}

To answer this question, we developed a preference prediction model with benchmark performance (ROC-AUC$=.630$ and F1-Score$=.711$) on user preference data collected across multiple grasp-object scenarios. The GraspR model supports both scenario-agnostic and scenario-specific configurations, enabling designers to balance generalizability and performance. We demonstrate the model’s applicability in four everyday scenarios where an adaptive grasp interface dynamically adjusts its layout for each task: baseball training, vegetable chopping, circuit board testing, and navigation. In summary, this work makes the following contributions:
\begin{enumerate}
    \item \textbf{GraspR}, the first computational model to predict spatial user preferences for adaptive and scalable grasp UI design.
    \item The first public \textbf{dataset} of user preferences in grasp interactions and a code repository to enable exploration and rapid design of adaptive grasp UIs.
    \item Data-driven analysis of \textbf{design factors} influencing user preference in grasp UIs.
    \item Working \textbf{prototype} of an adaptive grasp UI powered by GraspR.
\end{enumerate}

To facilitate future research and practical applications, we make our dataset and code publicly available\footnote{\url{https://github.com/HAL-UCSB/graspr}}.

\section {Usage Scenarios}
\label{sec:prototypes}

\begin{figure*}
    \centering
    \includegraphics[width=1\linewidth]{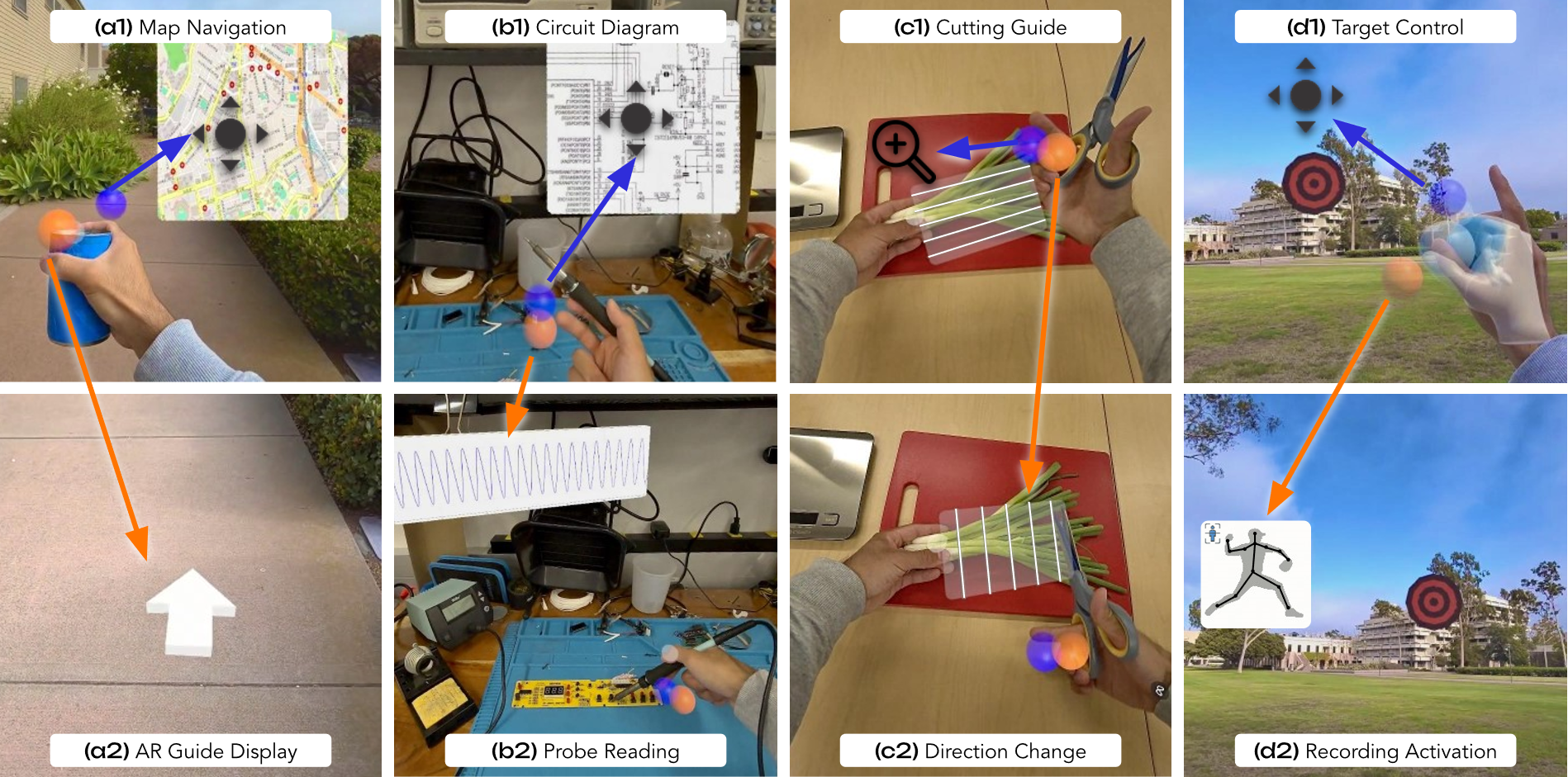}
    \caption{GraspR enables grasping interactions that adapt to changes in physical object grasp while maintaining alignment with user preferences. This new capability allows continuous use of grasp interfaces across a large variety of scenarios, such as: (a1) navigating a map with a joystick while holding a coffee; (a2) closing the map and activating and AR guide with a button; (b1) reading a circuit diagram with a joystick while holding a probe; (b2) starting a circuit measurement with a button; (c1) adjusting a cutting guide with a joystick while holding scissors; (c2) changing the direction of the cutting guide with a button; (d1) moving a virtual target with joystick while holding a baseball; (d2) starting body tracking recording with a button.}
    \label{fig:prototypes}
\end{figure*}

We first demonstrate the utility of GraspR through four scenarios that illustrate how it adapts UI elements to user preferences during single-finger microgestures—an interaction technique commonly explored in prior work~\cite{sharma2021solofinger, aponte2024grav}. As shown in Figure~\ref{fig:prototypes}, these examples highlight how GraspR enhances ergonomics, efficiency, and usability compared to static interfaces~\cite{ledo2018evaluation}. A prototypical implementation of GraspR was used to simulate these scenarios.

\begin{enumerate}

	\item Coffee-to-go Navigation:
		\begin{itemize}
		\item[$\bullet$] \textit{without GraspR}: A user holding a coffee cup interacts with an XR map interface. Fixed UI elements (e.g., zoom controls) require awkward finger stretches or grip adjustments, disrupting interaction flow and risking spills.

		\item[$\bullet$] \textit{with GraspR} (Figure~\ref{fig:prototypes}a): GraspR dynamically positions UI elements near the user’s fingertips, enabling seamless interaction without compromising their grip on the cup.

		\end{itemize}

	\item Circuit Testing on Electronics:
		\begin{itemize}
		\item[$\bullet$] \textit{without GraspR}: A technician uses an AR interface to navigate a schematic while holding a probe in one hand. Static controls force uncomfortable stretches,  probe repositioning, or the need to set down the probe, disrupting focus and increasing error risks.

		\item[$\bullet$] \textit{with GraspR} (Figure~\ref{fig:prototypes}b): GraspR adapts the placement of schematic navigation controls to the technician’s reachable, ergonomic zones, allowing them to interact with the interface naturally and without breaking their focus or probe position. This use case was inspired by Chatterjee et al.~\cite{chatterjee2022ardw}
		\end{itemize}

	\item Cooking with AR Guidance: 
		\begin{itemize}
        \item[$\bullet$] \textit{without GraspR}: A user is chopping vegetables in the kitchen while following an AR chopping aid. The tool’s interface controls (e.g., chopping size adjustment, toggle for the chopping marks) are fixed in positions that require the user to pause chopping and adjust their hand position to access them. This slows down the cooking process and leads to frustration.

		\item[$\bullet$] \textit{with GraspR} (Figure~\ref{fig:prototypes}c): GraspR predicts the user’s preferred interaction zones and places the cooking aid controls within natural reach, considering the user’s grasp on the scissors and the chopping motion. This allows the user to interact with the guide without stopping or adjusting their hand position. The chopping marks in this interface were inspired by Pohl et al.~\cite{pohl2024integrated}
		\end{itemize}

	\item Sports Training for Baseball Pitchers:
		\begin{itemize}
		\item[$\bullet$] \textit{without GraspR}: A pitcher using an XR training system must frequently adjust their grip on the baseball to interact with static interface elements, such as pitch trajectory visualization or target selection. These controls are fixed in non-optimal positions, disrupting the natural flow of the training session and detracting from realism.

		\item[$\bullet$] \textit{with GraspR} (Figure~\ref{fig:prototypes}d): GraspR predicts the pitcher’s natural hand movement and grip preferences and dynamically positions the training interface elements (e.g., visualization overlays, target selection) within easy reach. This enables the pitcher to interact seamlessly without disrupting their grip or breaking focus.
		\end{itemize}
\end{enumerate}

Manually adjusting UIs for varied grasp conditions is impractical, as is designing static interfaces for every scenario. GraspR automates this process, creating adaptive, preference-aware UIs that favor usability and ergonomics. By dynamically aligning UIs with user preferences, GraspR can potentially reduce strain, boost task efficiency, and enhance user experience across applications.

\section{Related Work}

Our work builds on grasp UI design---particularly computational approaches---as well as adaptive XR interfaces and user preference modeling in HCI.

\subsection{Grasp UI Design}

Many compelling use cases of grasp UIs are situated in the intersection of microinteraction and unimanual tasks~\cite{laviola20173d, guiard1987asymmetric}. Microinteractions are brief interactions with a secondary system that last at most four seconds and are designed to minimize interruptions to the primary task~\cite{oulasvirta2005interaction, ashbrook2010micro}. Examples of grasp UI use cases documented in the literature include music playback control while holding a bag handle~\cite{saponas2009enabling}, controlling a navigation system while grasping a steering wheel~\cite{doring2011gestural}, and activating text-to-speech features while holding a book~\cite{sharma2024graspui}. Input modalities for grasp UI include pressure gestures against a handheld object~\cite{saponas2009enabling, doring2011gestural, quinn2019active, villarreal2022theoretically} or motion gestures with single- or multi-finger movements---either in the air, on the surface of a handheld object, or the participants' hands~\cite{sharma2019grasping, sharma2021solofinger}. Purely gestural interfaces, without visual cues, have been criticized for their lack of discoverability~\cite{norman2010gestural}. To address this issue, prior work has proposed incorporating visual widgets to signify affordances of a grasp UI~\cite{ramos2004pressure, zeng2013thumb, zhou2020gripmarks, aponte2024grav, sharma2024graspui}. There are three main approaches to grasp UI design: Elicitation studies with end-users, expert-driven design, and computational design.

Elicitation studies, originally proposed by Wobbrock et al.~\cite{wobbrock2005maximizing}, are a common approach to grasp UI design. In an elicitation study, users propose symbols (e.g., gestures) for given referents (e.g., system actions). Designers then group similar symbols and assign each referent to the most frequently observed associated symbol to maximize guessability~\cite{wobbrock2005maximizing}. Saponas et al.~\cite{saponas2009enabling} conducted an elicitation study with twelve participants to create a muscle-activated grasp UI for controlling a music player while holding a mug or laptop bag. Villarreal et al.~\cite{villarreal2022theoretically} employed the method to propose 26 squeezing gestures for 21 smart home functions on a deformable cushion, such as pushes, wipes, and folds on different regions of the cushion. Sharma et al.~\cite{sharma2019grasping} identified a concise set of microgestures applicable across various scenarios through a large-scale elicitation study. They analyzed 2,400 microgestures to develop a compact taxonomy of 21 grasping gestures grouped into three grasp-type clusters: ``on-body,'' ``on-object,'' and ``in-air.'' These clusters inspired our gesture selection strategy as later presented. Although elicitation studies effectively incorporate user input into the design process of grasp UIs, they often require several hours of participants' time~\cite{sharma2019grasping}, which limits their scalability to a large number of grasp-object scenarios.

Expert-driven grasp UI design reduces over-reliance on user elicitation. This approach leverages generalizable taxonomies and guidelines, often combined with grasp behavior datasets, to inform design decisions~\cite{feix2015grasp, bullock2013grasp, corona2020ganhand}. For instance, Wolf et al. proposed a taxonomy of grasping microgestures based on interviews with ergonomics experts, considering use cases such as steering wheel control, stylus manipulation, and ATM card handling~\cite{wolf2011taxonomy}. Their taxonomy outlines the ergonomic feasibility and impact on primary tasks across various grasping gestures, highlighting how scenario-based constraints affect the usability of such interfaces. Joshi et al. proposed a design space of transferable microgestures~\cite{joshi2023transferable} comprised of constraints and transferability dimensions. In their framework, grasping hands are considered the most restrictive constraint for gesture design, but also present rich examples of gestures integrated with tangible objects. Zhang et al.~\cite{zhang2024stick} developed a comprehensive taxonomy of real-world affordances of stick-shaped objects accompanied by a gesture set. Sharma et al.~\cite{sharma2021solofinger} examined the detection robustness of single-finger grasping microgestures. By analyzing an existing kinematic dataset of daily activities and machinery tasks~\cite{bullock2013grasp}, they showed that single-finger movements are rare. They further demonstrated that single-finger grasping microgestures can be reliably detected using hand-tracking gloves. Building on these findings, we focus on single-finger motion as the preference items in our model. It is not rare to find instances of grasp UI designs that combine expert-driven design and elicitation studies to strike a balance between designer goals and users' perspectives~\cite{saponas2009enabling, sharma2019grasping, villarreal2022theoretically}. While resources for expert-driven grasp UI design have multiplied, this approach cannot support adaptations to grasp-object scenarios beyond those anticipated during the design phase.

Computational design methods for grasp UIs enable adaptive designs even in scenarios unforeseen by designers. For example, in GraV~\cite{aponte2024grav}, researchers developed a forward kinematics simulator for single-finger movement constrained by grasp, considering the joints' range of motion, and hand and object shapes. The resulting point clouds encoded joint angle distance values, which prompted workshop participants to change their grasp UI designs. Zhao et al.~\cite{zhao2024grip} developed a simulation tool to evaluate the trade-off between grasp stability and object surface reachability for grasp interactions. The tool uses 3D models of hands and objects to simulate grasp quality, which measures how securely the object can be held, and reachability, which assesses how much of the object's surface can be accessed by fingers for interaction. Although scalable, the lack of user preference information in current computational tools for grasp UI design can lead to adaptive designs that are geometrically and anatomically feasible but undesired by users.

GraspR addresses this gap by enabling scalable computational design of grasp UIs informed by user preferences. Additionally, we contribute to expert-driven approaches for grasp UI design with an analysis of how design factors influence user preferences. To accelerate research in this domain, we share our dataset containing 1,520 data points of user preferences in grasp interactions. Finally, our proof-of-concept demonstrates how adaptive grasp UIs can be effectively implemented in real-world scenarios.

\subsection{Adaptive Interfaces in XR}

Adaptive User Interfaces (AUIs) can dynamically alter their structure and behavior to remain functional across context use variations~\cite{greenberg1985adaptive, coutaz2010user}, as opposed to static interfaces that remain unchanged throughout an application’s lifespan. Interface adaptations often respond to changes in the context by updating their features, layouts, widgets, information structure, or content density~\cite{akiki2014adaptive}. Notably, Gajos et al. proposed ARNAULD~\cite{gajos2005preference}, a system that adapts 2D interfaces based on cost functions learned from user preferences. Inspired by their approach, GraspR enables adaptive grasp UIs.

Adaptability is essential in XR interfaces, as many constraints impacting these interfaces are hard to anticipate at design time. E.g., geometry of the surroundings, semantics, social dynamics, and user motor abilities. Lindlbauer et al.~\cite{lindlbauer2019context} proposed a method to automatically adapt the level of detail in response to changes in users' estimated cognitive load. AUIT~\cite{evangelista2022auit} is an adaptive XR interface toolkit that enables adaptions through multi-object optimization, such as visibility and reachability, while allowing customizations. Closely related to our work, XRgonomics~\cite{evangelista2021xrgonomics} offers a spatial model of ergonomic cost for mid-air interactions, supporting ergonomic adaptive XR interfaces. GraspR contributes to adaptive XR interfaces with a computational model of user preference in grasp interactions that informs grasp UI adaptations, enhancing XR integration with objects in physical-virtual dual-task scenarios.

As Mixed Reality (MR) becomes more socially prevalent, adaptive interfaces must consider the social aspects of shared spaces. SituationAdapt~\cite{li2024situationadapt} combined an optimization engine inspired by AUIT with a Vision-and-Language Model (VLM) to reason about social contexts and dynamically adjust MR interfaces. GraspR complements this vision of socially integrated MR by enabling adaptive grasp UIs that support subtle interactions, potentially offering more socially acceptable alternatives to mid-air gestures and controller-based interactions.

AdapTUI is a framework that adapts geometric-feature-based tangible user interfaces across different physical environments by leveraging similar geometric shapes (e.g., edges, planes)~\cite{he2024adaptui}. This optimization approach addresses geometric consistency, ergonomics, visibility, and spatial-temporal consistency. However, user feedback revealed failure cases where adapted layouts were incompatible with user preferences. GraspR conciliates user preference and spatial features by modeling the relationship between grasp-based interactions.

\subsection{User Preference Modeling}

User preference in HCI is considered from both formative and summative perspectives. Researchers seek to identify design factors that enhance user preference, which is also used as a key criterion for comparing different interface alternatives. User preference is known to be influenced by multiple factors such as performance~\cite{nielsen1994measuring}, aesthetic qualities~\cite{tractinsky2000beautiful, koyama2016computational}, and trade-offs of rewards and ease of use~\cite{toomim2011utility}. When presented with alternatives, participants inform their preferences not only explicitly but also through non-verbal cues such as facial expressions~\cite{tkalvcivc2019prediction} or EEG signals~\cite{zhang2024identifying}. Symbolic elicitation studies have also been considered a method to unveil participants' unconstrained preferences for generating symbol-referent associations in tune with participants' mental models~\cite{vatavu2016between}, which is captured by an agreement coefficient~\cite{vatavu2022clarifying}.

In Decision Theory, preference refers to how a rational agent compares a set of options based on expected satisfaction~\cite{steele2015decision}. Preference can be formalized as a binary relation on a set of alternatives defining an order among those options. In ordinal preferences, only the rank of the alternatives matters. Conversely, cardinal preferences do consider the magnitude of the differences between items. Revealed Preference Theory~\cite{samuelson1938note} holds that individual preferences can be inferred from observed choices, without relying on introspective or hypothetical statements about preferences. According to the Utility Representation Theorem~\cite{debreu1954representation}, if a preference relation is complete, transitive, and continuous, then there exists a utility function that assigns a real number to each option such that option A is strictly preferred to B if and only if the utility of A is greater than that of B, thus preserving the agent’s preference ordering. Informed by these theories, we developed GraspR as a utility function learned from observed participants' choices between alternative single-finger grasp interactions, such that the predicted probability that one option is preferred over another can be interpreted as a cardinal preference.

\section{GraspR Preference Dataset}

To develop a computational model of user preferences in grasp interactions, we gathered empirical data in controlled conditions across grasp-object combinations. Although existing datasets of grasp interactions are valuable resources, they lack user preference annotations~\cite{sharma2021solofinger, aponte2024grav, sharma2024automated}. To address this gap and enable preference-aware adaptation in this domain, we collected the GraspR dataset of user preferences in grasp interactions.

\subsection{Data Collection Methodology}
Our data collection process followed a 2-alternative forced choice (2AFC) paradigm, where users are presented with two distinct alternatives and must select one. This experimental design is common in perceptual threshold studies and user preference elicitation~\cite{feick2024predicting, zhong2021spacewalker, gajos2005preference}. This method is especially suitable for our goals as it eliminates neutral responses, reduces central tendency bias, and yields clearer, more decisive preference data.

To contextualize the quantitative results, we ran semi‑structured interviews using simulated grasp UI scenarios~\cite{oulasvirta2003understanding}. This approach added qualitative data to our understanding of user preferences in grasp interactions, which can vary in time and across tasks.

\subsubsection{Participants}

We recruited eight participants: four females (two left-handed and two right-handed) and four males (one left-handed and three right-handed). Our participant demographics align with prior grasp interaction studies~\cite{fan2023arctic, bullock2013grasp}. The participants' ages ranged from 18 to 42 years. Seven of them had used a MR headset before. None of the participants had a history of dominant hand surgery, fracture, amputation, arthritis, or other musculoskeletal conditions, maintaining typical hand function across the sample. Before the data collection, we measured participant hand lengths (wrist to middle fingertip: $M=18.16cm,~SD=1.8cm$) and metacarpal breadth (index to little finger bases: $M=7.81cm,~SD=0.89cm$)
~\cite{chao1989biomech}. Hand metrics were within typical ranges of the adult population reported in other studies and human factor guides~\cite{dodhfe2000human, bergstrom2014modeling}. Sessions lasted one hour and participants were compensated at a rate of US\$20/h. All participants provided informed consent, and our local IRB approved this study \emph{(protocol number 22-23-0317)}. Figure~\ref{fig:setup}e shows how participants are grouped in our data collection.

\subsubsection{Grasp-Object Selection}
\label{sec:grasp_objects}

Our data collection involved four distinct grasp-object conditions, as shown in Figure~\ref{fig:setup}a. We selected the grasp types in our data collection by ranking the 33 grasp types studied by Bullock et al.~\cite{bullock2013grasp} by the product of their use frequency and duration, highest to lowest. From this ranking, we selected the four highest-ranked grasps that collectively covered all levels across the three GRASP taxonomy~\cite{feix2015grasp} dimensions---power-precision spectrum, thumb positions, and opposition types---ensuring at least one instance in each taxonomic category. Together, the selected grasp types account for 48\% of the total frequency-duration product reported by Bullock et al.~\cite{bullock2013grasp}, forming a compact and yet representative set of grasp types.

We identified objects associated with our selected grasps using the YCB Affordances dataset~\cite{corona2020ganhand}, which maps objects from the standardized YCB object set~\cite{calli2015ycb}, widely used in grasp research, to Feix et al.'s GRASP taxonomy~\cite{feix2015grasp}. The chosen objects range from simple geometries to complex articulated forms, all related to real-world activities. This systematic approach ensured a wide coverage of fundamental grasp variations and object geometries. We covered all sharp edges and pointed tips to ensure participant safety. The final selected grasp-object conditions were:
\begin{enumerate}
    \item \textbf{Medium wrap (soup can)}: A thumb-abducted grasp and a basic cylindrical shape typical in cups and medium-radius cylinders.
    \item \textbf{Lateral tripod (racquetball)}: a thumb-adducted grasp and a basic sphere.
    \item \textbf{Writing tripod (whiteboard marker)}: a grasp for dexterous tasks and a thin cylinder common in precision tools and writing instruments.
    \item \textbf{Distal (scissors)}: power grasp with pad opposition and an articulated object that constrains fingers.
    
\end{enumerate}


\subsubsection{Grasp UI Gestures}
\label{sec:targets}

\begin{figure}
    \centering
    \includegraphics[width=.9\linewidth]{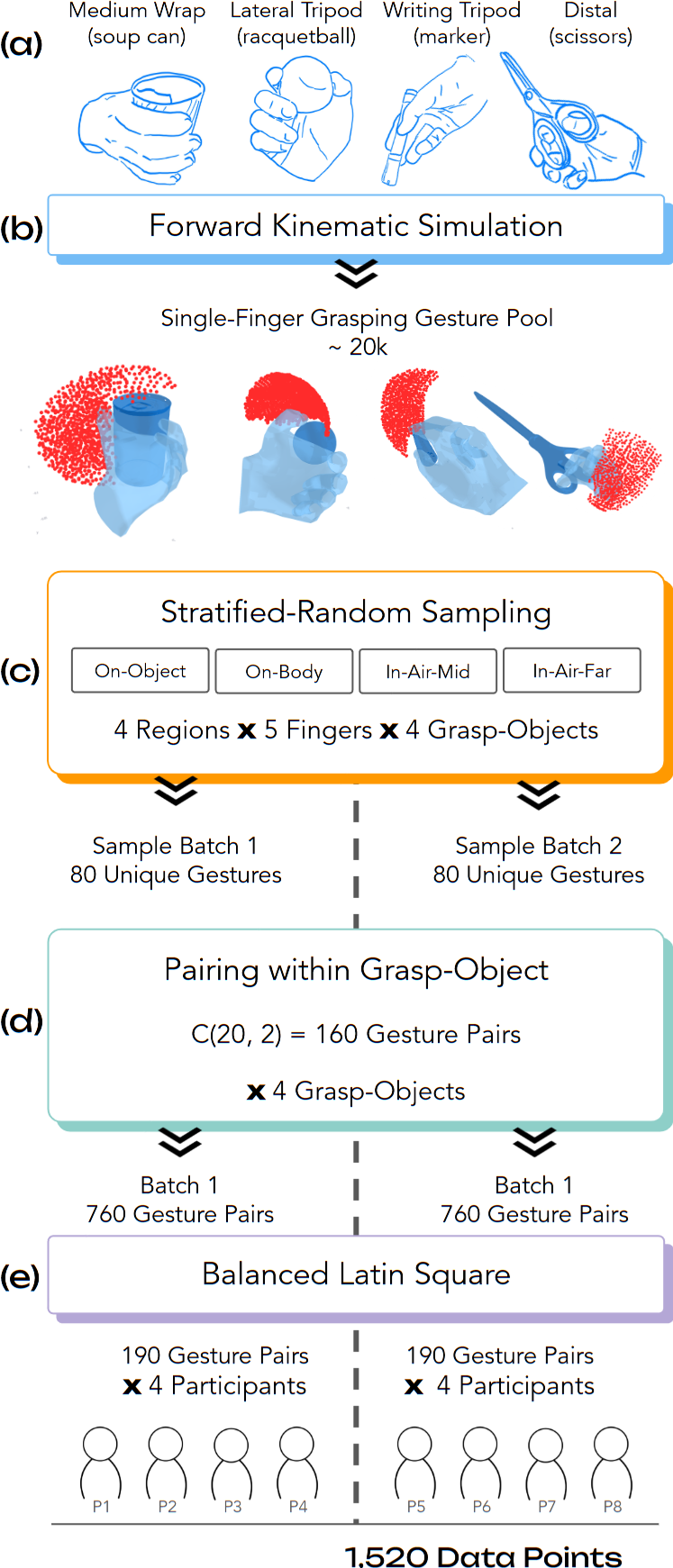}
    \caption{Gesture pairs compared by participants: (a) The four grasp-object conditions in our data collection span the main dimensions of the GRASP taxonomy~\cite{feix2015grasp} and include objects with both simple regular shapes and complex articulated structures. (b) A forward kinematic simulator~\cite{aponte2024grav} generated a large pool of single-finger grasping gestures for each condition. (c) We select two batches from the gestures pool stratified by fingers and four interaction regions. (d) We paired gestures within the same batch and grasp-object stratum. (e) Participants compared gesture pairs in a counterbalanced grasp-object order determined with a balanced Latin square.}
    \label{fig:setup}
\end{figure}

After choosing the four object-grasp conditions, we generated a large pool of gestures and selected gesture pairs for participants to compare during each grasping task. We focused on the single-finger reach gesture, defined as the movement of a fingertip from its initial grasp position to a target point and back. This gesture mirrors the point-and-click interaction in cursor-based interfaces, making it an intuitive and fundamental component of grasp interactions. This basic movement can combine to form more complex gestures, such as taps, swipes, flexes, and extensions, including microgestures proposed by Sharma et al.~\cite{sharma2021solofinger}.

We chose single-finger interactions for three key reasons, building on prior work that has explored grasp-constrained or unimanual interactions in XR~\cite{aponte2024grav, pfeuffer2023palmgazer, sharma2024graspui}. First, in many real-world grasp scenarios, the hand is partially or fully occupied, leaving only one or two fingers available for interaction. Second, modeling single-finger input enables precise analysis of spatial comfort and reachability, isolating user preferences without interference from multi-finger coordination. Third, single-finger gestures align with common AR input methods, such as pointing or tapping, providing a foundation for more complex grasp-based UI interactions.

In our study, we define each gesture as a tuple of finger label, initial fingertip position on a grasp, and target point. To ensure realistic variations in the comparison pairs, we generated approximately 20,000 single-finger gestures using a forward kinematic simulation method proposed in prior work~\cite{aponte2024grav}. The grasping gesture simulator consists of a flood-fill algorithm on the range of motion of each finger, limited by object and hand geometry. Each step of the simulation stores the joint rotations of hand pose in a feasible single-finger gesture. We applied this method to all five fingers across four grasp-object conditions. To further ensure reachability, we discarded gestures towards the farthest 10\% targets. Figure~\ref{fig:setup}b shows reachable points from single-finger motions during grasping of each object, generated via forward kinematic simulation. 

\begin{figure}
    \centering
    \includegraphics[width=.9\linewidth]{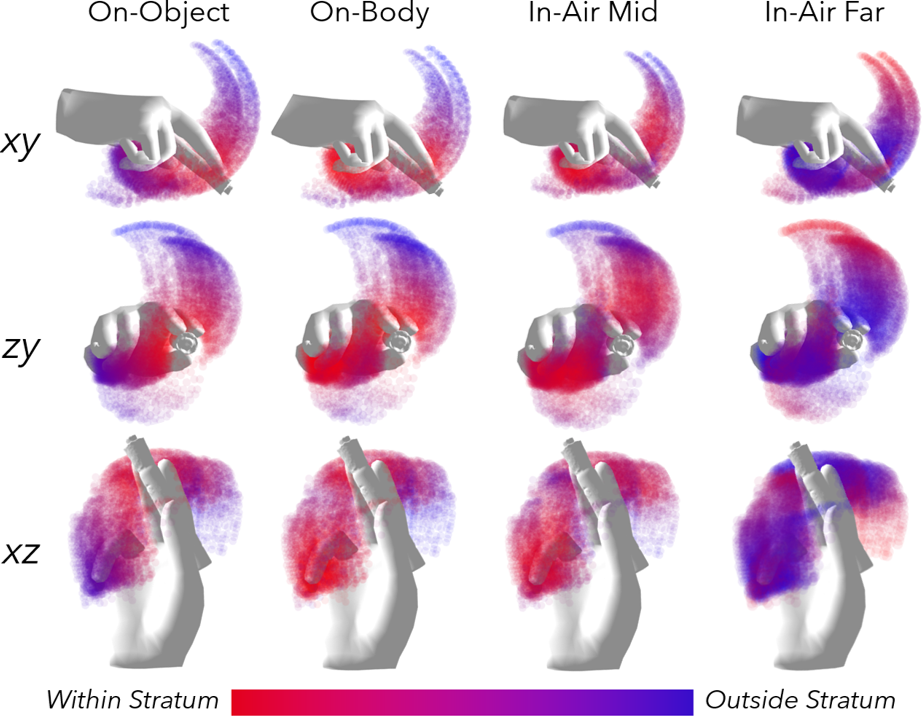}
    \caption{We segmented the simulated gesture pool into four regions: on-object, on-body, in-air close, and in-air far. Within each region, we selected representative single-finger gestures, prioritizing those with the closest target points.}
    \label{fig:strata}
\end{figure}

We segmented the generated in four strata inspired by the regions in Sharma et al.'s taxonomy~\cite{sharma2019grasping}: ``on-object,'' ``on-body,'' and ``in-air,'' further divided into ``in-air-mid'' and ``in-air-far.'' Each stratum is procedurally defined by a distance function. Target points near the object’s surface form the on-object strata. Those closest to the stationary part of the hand (e.g., the palm or stationary fingers) form the on-body strata. Target points near the medoid of the space reachable by the finger belong to the in-air-mid stratum. Lastly, the in-air-far stratum consists of the farthest points from the initial position. Figure~\ref{fig:strata} shows an example of the resulting segmentation. To ensure balanced coverage of single-finger gestures, we cycle through these strata in a round-robin fashion and, without replacement, select the point with the smallest distance in each stratum. This approach ensures that selected gestures are balanced across all four regions, preventing any single stratum from dominating the dataset. Figure~\ref{fig:setup}c illustrates the stratified selection of gestures in the data collection setup.

Once the gestures were defined, the next step was to combine them in pairs so participants could compare them in the data collection task. To achieve a balanced set of gesture alternatives while avoiding a combinatorial explosion, we paired gestures across different fingers and strata but only within the same batch and grasp-object condition. By avoiding pairs that cross between different objects, we eliminated potential object-specific biases and minimized object switching time during data collection, optimizing participant time and effort. Figure~\ref{fig:setup}d shows how we assigned gesture pairs to participants in batches.

\subsubsection{Apparatus}
\begin{figure*}
    \centering
    \includegraphics[width=\linewidth]{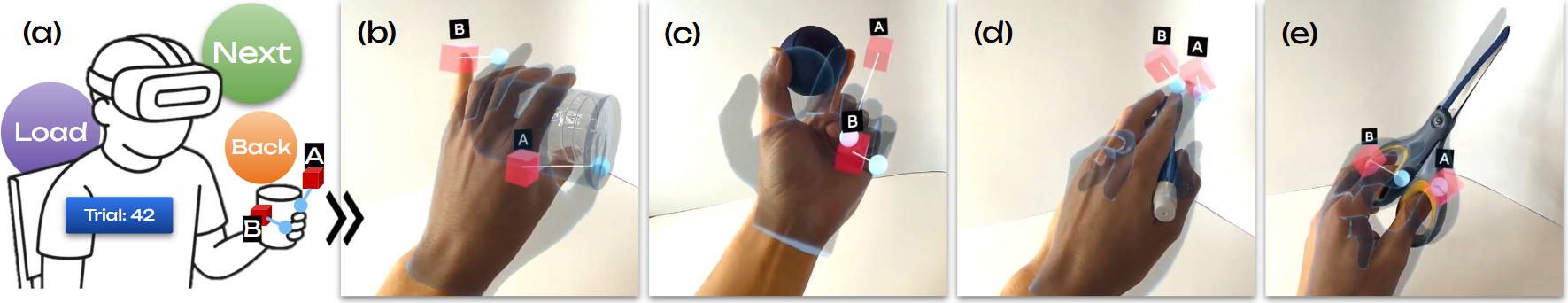}
    \caption{Study setup: (a) Mixed-reality controls allowed participants to load and navigate back and forward through trials. (b-e) In each trial, participants expressed their preference between two single-finger gestures performed while holding an object. The interface guided participants' gestures with lines connecting a blue sphere in their fingertips to the target red cube. Each gesture was labeled as ``A'' or ``B.'' The same visualization was used for all the pairwise combinations across the grasp-object conditions: (b) Medium wrap on a can; (c) Lateral tripod on a racquetball; (d) Writing tripod on a marker; (e) Distal on scissors.}
    \label{fig:static_viz}
\end{figure*}

We used a Meta Quest 3 headset to display the interface during data collection. To guide participants toward the correct grasp pose, the interface displayed a static 3D model of a hand holding an object. This visual reference is shown as the dark overlay in Figure~\ref{fig:static_viz}. This model was anchored to participants' dominant wrist and scaled to match their hand measurements. The 3D models of both the hand grasp and the object were obtained from the YCB Affordance dataset~\cite{corona2020ganhand}. Located next to the person, as shown in Figure~\ref{fig:static_viz}a, are three spatial buttons: ``Load'' to load the gesture pair for comparison, ``Next'' to move to the next pair of target points, and ``Back'' to go back to the previous pair. A sequence number was displayed in the user's field of view.

In our MR interface, grasp UI gestures were visualized as a line connecting a blue sphere (on the fingertip) to a red target cube. Each target was a $1.5~cm^{3}$ red cube labeled as either ``A'' or ``B,'' consistently sized across all trials. This size provides a balance between visibility and the need for fine motor control. It is slightly larger than the $1~cm^{2}$ recommendation for interactive touch targets \cite{nngroup_touch_targets}. Although spheres are more aligned with Fitts' Law analysis, our pilot studies found cubes provided better depth perception. We used identical visual representations for all target points to prevent bias toward a particular visualization style. 

We utilized a static visual representation instead of a dynamic interactive interface for two reasons. First, our study focused on collecting subjective preferences between single-finger gestures rather than evaluating usability. Second, this approach reduced visual load and avoided bias toward areas better detected by current commercial hardware~\cite{wobbrock2009user}.

\subsubsection{Procedure}

We assigned participants to a sequence of grasp-object conditions counterbalanced with a balanced Latin square~\cite{schwind2023hci}. The experimenter explained the procedure to the participants and emphasized that they did not need to justify their choices. This is an important instruction, as the expectation of justification can increase decision difficulty, promote compromise choices, and discourage the use of hard-to-explain criteria~\cite{lerner1999accounting}.

For each 2AFC trial, the MR interface instructed participants to grasp the object and displayed two gestures simultaneously. We used participants' hand measurements to determine a scale factor between their hands and the grasp visual reference. This scaling ensured all target points remained reachable regardless of hand size. The experimenter then asked, ``Which one do you prefer?'' and encouraged participants to repeat each gesture as many times as needed, with no strict time limit. After deciding, participants verbally reported their preference, which the experimenter directly recorded in a control spreadsheet. Neutral responses were not allowed. Participants then touched the button ``Next'' to move on to the next pair. The preference labels in our dataset are $1$ if the participant selected option ``A'' and $0$ otherwise. Throughout the session, the experimenter monitored participants through the headset video stream to ensure they followed the procedure and adjusted their grasp when necessary. The experimenter could check whether participants skipped a pair by mistake by checking a sequence control number. If that happened, participants were requested to click the ``Back'' button until the sequence numbers matched.

After the 2AFC trials, we conducted a 10-minute semi-structured interview with participants to explore contextual factors influencing grasp UI preferences that were not part of the 2AFC data collection. The interviews began with general questions about the 2AFC trials: \textit{``What object was the hardest to interact with?''} and \textit{``What finger was easier to use?''} Next, the interviewer presented a hypothetical MR media player system and asked participants to simulate everyday activities like drinking coffee, eating, or writing. Within these scenarios, participants were asked \textit{``How would you skip N songs?''} and given alternatives such as using a mid-air MR button (represented by a paper sketch held by the interviewer) or any interaction method from the 2AFC study. Inspired by the bodystorming~\cite{oulasvirta2003understanding} technique, we encouraged participants to physically perform these gestures, allowing them to set down objects or use both hands if they did so. For example, some participants were asked to simultaneously hold an apple (simulating a lateral tripod grip on a sphere) and a cup of juice (simulating a medium wrap on a cylinder).  We varied scenarios by changing object properties (e.g., temperature, weight) and increasing the number of songs to skip, to understand the perception of effort during repeated use.

\subsection{Dataset Summary and Preprocessing}

We collected a total of 1,520 preference-based choices between single-finger grasp interaction alternatives using a 2AFC method. Each entry in the dataset records the user's binary preference, labeled as 1 or 0. To complement the quantitative data, we conducted semi-structured interviews that captured user reflections on contextual factors influencing their choices. These qualitative insights provide a holistic view of user preferences in grasping UI. 

For preprocessing, we extracted a set of features from each trial, including finger and object positions, angles, and volumetric measures (see Section~\ref{sec:factors}). These features were calculated based on the simulation outputs and were selected to capture spatial and ergonomic characteristics of each grasp gesture. We then applied correlation analysis, variance analysis, and logistic regression to examine how these factors relate to user preferences. We interpret these features through their odds ratios. 

Our sensitivity analysis for logistic regression\footnote{G*Power 3.1.9.7, two tails, $\alpha=.05$,  $1-\beta=.8$, $N=1,520$, uniform X} showed that the dataset size (1,520 samples) is sufficient to reliably detect odds ratios of at least 1.68 (68\% increase in odds) and at most 0.55 (45\% decrease in odds), both below the conventional medium effect size threshold. Additionally, a priori power analysis for logistic regression indicates that a sample size of 486 would be sufficient to reliably detect a medium effect size\footnote{G*Power 3.1.9.7, two tails, $\alpha=.05$,  $1-\beta=.8$, odds ratio=2.477, uniform X}.

\section{GraspR Design Factors}\label{sec:factors}

We hypothesize that differences in their underlying features can explain user preferences between grasp UI alternatives. For each pair of interface options presented to users, we calculate the feature difference as $\Delta_{feat}=A_{feat} - B_{feat}$, where $A_{feat}$ denotes the feature values from alternative ``A'' seen in Figure~\ref{fig:static_viz}(b-e). These features are calculated based on data produced by the simulation process that generated the single-finger grasping gestures used in our data collection. We now present these features and an analysis of their relationship with empirical user preference in grasp UIs. Table~\ref{tab:factors} summarizes our findings.

\subsection{Features of Grasp Interactions}
\label{sec:feats}

\begin{figure}
    \centering
    \includegraphics[width=.8\linewidth]{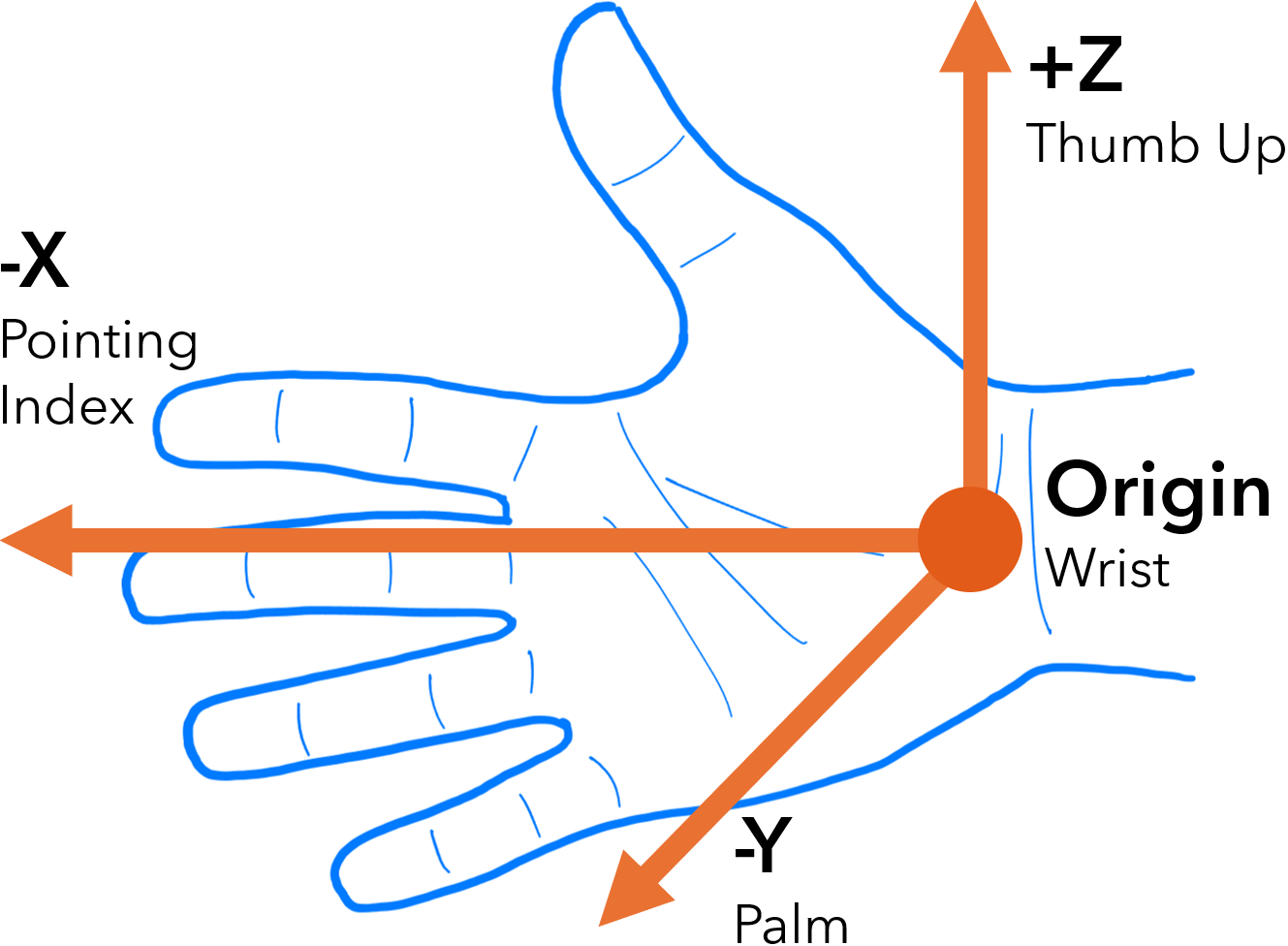}
    \caption{GraspR uses a wrist-centered coordinate system. The Z-axis points toward the right thumb, the Y-axis opposite the palm, and the X-axis opposite the index finger.}
    \label{fig:hand_coords}
\end{figure}

\paragraph{Positions \& Angles}
Our simulation captured fingertip positions, joint rotations, and 3D meshes of both the object and static hand components (palm and non-moving fingers) at each step. From this data, we calculated the following feature differences between gesture alternatives: target positions, initial fingertip position, distance from the initial fingertip position to the target, the smallest distance from the static parts of the hand to the target, and the smallest distance from the object to the target. Inspired by the cost metric utilized by Aponte et al.~\cite{aponte2024grav}, we compute the sum of quaternion distances between joint angles and the sum of Euler angle distances. We adopt a coordinate system centered at the wrist joint, as depicted in Figure\ref{fig:hand_coords}.

\paragraph{Volumetric}
Over time, the movement range of the fingertips forms a volume bounded by finger mobility, the object’s surface, and the hand itself~\cite{aponte2024grav}. This reachable space determines regions for interactivity, e.g., pressing against the surface of the object, sliding on the body, or flicking a MR button surrounding the grasp. After the simulations are complete and the entire reachable space is known, we can calculate the differences in finger reachable volume, the smallest distance from the target to the convex hull of finger reachable space, and the ratio of finger reachable volume and all fingers' reachable volume. Inspired by Zhao et al.’s grasp quality assessment~\cite{zhao2024grip}, but differing in approach, we adopted a purely geometric metric well-established in robotics and ergonomics: the grasp cage ratio~\cite{vahedi2009complexity}. This metric carries no physical assumptions and is defined as the ratio between the volume of the convex hull formed by hand–object contact points and the volume of the object’s convex hull. Intuitively, values closer to one indicate a more stable grasp with fewer escape routes, while lower values suggest weaker grasp support. To compute the grasp cage, we approximated the contact surface using the vertices of the object’s colliding faces as contact points. Gesture alternatives are compared with the difference in cage ratios.

\paragraph{Guidelines}
Design guidelines from previous research might help to understand user preferences in grasp UIs. In particular, we include the average finger individuation index for grasping hands~\cite{sharma2021solofinger} and free hands~\cite{hager2000quantifying}. This metric ranges from $0$ to $1$ and is larger for fingers that can move more independently from the others (e.g., $IID_{thumb}=.98>IID_{ring}=.95$). Our feature set also includes anatomical loss proposed by Yang et al.~\cite{yang2021cpf}, which penalizes abnormal hand poses when fitting MANO (hand Model with Articulated and Non-rigid defOrmations)~\cite{romero2017embodied} models.

\subsection{Feature Analysis}

To inform adaptive grasp UI design, we seek to identify which design factors measurably influence user preferences. We conducted quantitative analyses to uncover statistically significant relationships between various design factors and user-preferred single-finger gestures. Quantifying the relationship between these features and user preference offers concrete guidelines for expert-driven design of grasp user interfaces. By identifying which features significantly influence user preference, designers can make informed decisions even without access to computational design tools or direct end-user feedback. This evidence-based approach allows designers to anticipate user preferences and optimize interfaces based on empirical data rather than intuition alone. For this reason, we seek a minimal feature set of statistically significant features related to user preference in grasp UI design.

Since our goal with this feature analysis is to provide concrete design guidelines, we conduct our analysis on the entire dataset. To avoid data leakage and misleading evaluations, the resulting feature set will not be used for training GraspR. The feature selection, training, and evaluation processes for GraspR are presented independently in Section~\ref{sec:model}.

In the first step of our quantitative analysis, we employed class entropy to verify that the preference labels in our dataset are balanced ($a=56.16\%,B=43.84\%,H=.98$). Note that preferences for the ``A'' alternative are encoded as $1$ and as $0$ otherwise. We then proceeded to inspect the relationship between the differences in features between the alternatives and the preference label. The differences in features of the single-finger grasp alternatives were calculated as $\Delta_{feat}=A_{feat} - B_{feat}$, where $A_{feat}$ denotes features from the alternative labeled with ``A'' as seen in Figure~\ref{fig:static_viz}. 

The differences in the ranking of finger ease of use exhibited the strongest positive correlations with the preference label ($r_{pb}=.18, r_{s}=.17,p<.001$). The difference in distances from the initial fingertip position to the target showed the strongest negative point-biserial and Spearman's correlations with the preference label ($r_{pb}=-.22,r_{s}=-.21, p < .001$). However, all of them are considered weak correlations.
The difference in ratio between finger-reachable volume and the entire hand-reachable volume had the highest mutual information score ($I=.054$), followed closely by the difference in distances from the initial fingertip position to the target ($I=.049$).

To identify a minimal feature set significantly related to user preference, we reduced collinearity by iteratively removing features with the highest variance inflation factors until all remaining features fell below our threshold ($VIF<10$).
We trained a logistic regression with L1 regularization (LASSO, $\lambda=1$) to iteratively eliminate features with coefficients $\beta=0$.
From the remaining features, we maintained only the ones with significant $\beta$ ($p<.05$).

Using the $\beta$ coefficients from this logistic regression, we can estimate how a one-standard-deviation change in feature differences affects the odds of user preference. The analysis of the odds ratio of the final selected features revealed that single-finger grasping gestures performed with fingers with a high individuation index (e.g., thumb and index) are preferable. More precisely, an increase of one standard deviation in the differences of grasping finger individuation index increases the odds of user preference by 38\%. Another factor that contributes positively to user preferences is the distance from the target to the limits of the reach space. Regarding the odds ratio, an increase of one standard deviation on the target distance to the reach limit increases the odds of preference by 13\%. Therefore, \textbf{we recommend designers to favor single-finger grasping interactions with the thumb and index finger towards the center of the reachable space of those fingers.}

Conversely, participants did not prefer single-finger grasping gestures with target points either too close to the wrist or too close to the palm. An increase of one standard deviation in the Y-coordinate of the target decreased the odds of preference by 36\%. Similarly, an increase of one standard deviation in the X-coordinate of the target decreases the odds of preference by 37\%. Therefore, \textbf{we recommend designers to avoid single-finger grasping interactions that require flexing the fingers towards the wrist.}

\begin{table*}
    \centering
    \caption{Impact of the significant features on the user preference odds ordered by impact in odds of preference. Participants preferred targets accessible with small movements of the thumb and index fingers. $\Delta$ Feature denotes the difference in feature values between alternatives ``A'' and alternatives ``B'' compared by participants in the 2AFC trials. $\beta$ logistic regression coefficients denote the change in the log-odds of preference for a one-unit increase in the respective feature.}
    
    \begin{tabular}{lccc}
    \toprule
    $\Delta$ Feature & $\beta$ & p-value & Impact in Odds \\
    \midrule
    Finger Individuation Index & .325 & <.001 & +38\% \\
    Distance from the Target to Reach Limit & .128 & .004 & +13\% \\
    \hline
    Distance from the Target to Body & -.236 & .001 & -21\% \\
    Reachable Volume & -.267 & <.001 & -23\% \\
    Anatomy Loss & -.269 & <.001 & -23\% \\
    Distance to Target & -.330 & .012 & -28\% \\
    Target Y-Coordinate & -.447 & <.001 & -36\% \\
    Target X-Coordinate & -.471 & <.001 & -37\% \\
    \bottomrule
    \end{tabular}
    
    \label{tab:factors}
\end{table*}

\section{GraspR Model}
\label{sec:model}

Our approach to evaluate GraspR consists of comparing two complementary classifiers: Logistic Regression (with L2 regularization, $\lambda=2$, SAGA optimizer) and Random Forest (50 estimators). This selection balances interpretability through Logistic Regression's transparent feature coefficients with the predictive power of Random Forest's ability to capture non-linear relationships. 

To ensure robust evaluation across different users, we implement Leave-One-Subject-Out Cross-Validation (LOSOCV), where each participant's data serves as a test set for a model trained on all other participants' data. This validation approach prevents participant-specific data leakage and better estimates model generalizability to new users. The reported results are averages of all the LOSOCV folds. We evaluate ROC-AUC, which measures overall classification quality, and F1-Score at the $.5$ threshold, which balances precision and recall in a single metric.

\subsection{Generalist Model}

The Logistic Regression achieved both better discriminative ability (ROC-AUC$=.630$) and superior precision-recall balance (F1-Score$=.711$). Table~\ref{tab:eval} category ``Generalist'' shows ROC-AUC and F1-Score for both Logistic Regression and Random Forest estimators. With these results, GraspR establishes the first benchmark for the challenging task of predicting user preferences across highly variable grasp interaction conditions. Even with simple estimators and minimal tuning, GraspR demonstrated predictive power that can help designers and researchers achieve adaptiveness at a large scale, saving countless hours of design and testing.

\subsection{Expert Models}

GraspR serves as a general model of user preferences in grasp UIs, agnostic to users, fingers, and object-grasp conditions. However, a practical grasp UI design might focus on interactions within more specific contexts. For instance, designers might prioritize adaptive grasp interactions for specific fingers (e.g., the little finger) while users hold particular objects (e.g., mugs or umbrellas). In such cases, designers may prefer to trade off generalizability for enhanced performance within their relevant constraints. To address these needs, we evaluate expert versions of GraspR that are optimized for higher performance in these targeted scenarios.

In a grasp-object specific configuration, the logistic regression obtained higher ROC-AUC scores than the random forest, while the random forest achieved the best F1-Scores. 
The GraspR trained specifically on the lateral tripod over racquetball had a boost of $.060$ in F1-Score and $.061$ ROC-AUC over the generic version. This increment offers designers a performance benefit at the expense of a less generalizable model, suitable for constrained design scenarios where designers only care about a specific grasp-object condition. Similarly, in the finger-specific version of GraspR, the highest boost was observed in the thumb-specific GraspR version, $.089$ in F1-Score. Finger-specific GraspR was trained on partitions of only 48 samples, which contained pairs of gestures performed with the same finger. When considering region-specific GraspR versions, increases were more modest, of at most $.013$ in ROC-AUC for the in-air-mid region.

Not all specialized versions of GraspR outperformed the generic model. In several cases, particularly the middle- and ring-finger-specific versions, the benefits of reduced variability were outweighed by the limitations of smaller training datasets.

\subsection{Data Ablation}

To assess model generalizability, we conducted a data ablation on grasp-object conditions using a leave-one-subject-out approach. As expected, ROC-AUC dropped in distal-scissors (LR$=.535, $RF$=.529$), writing-tripod-marker (LR$=.595$, RF$=.539$), and lateral-tripod-ball (LR$=.579$, RF$=.552$). The performance drop relative to the generalist model shows that each fold is nonredundant and contributes uniquely to predicting broader preferences. In contrast, ablating medium-wrap-can yields a modest ROC-AUC increase (LR$=.641$, RF$=.627$) over the generalist model. These gains, along with consistent improvements in the specialist model, suggest a domain difference for this fold relative to the rest of the dataset.

Relating this to our qualitative data, participants described the can as the ``easiest'' and most stable grasp, reinforcing its distinctiveness. We recommend the use of the specialist model for medium-wrap-can (LR$=.71$, RF$=.63$) and the generalist model for the other conditions. Similar recommendations have been made in the handheld touchscreen devices~\cite{buschek2013user}. Medium-wrap is the most common grasp reported by Bullock et al.~\cite{bullock2013grasp}, accounting for 40\% of total usage frequency-duration. Thus, performance gains in this condition may have greater relevance for real-world deployments. Table~\ref{tab:ablation} summarizes the results of our ablation study.

\begin{table}[]
\caption{Performance evaluation of expert vs. generalist versions of GraspR. The generalist model is trained on data aggregated across all categories.}
\label{tab:eval}
\begin{tabular}{@{}lccccl@{}}
 & \multicolumn{2}{c}{ROC-AUC} & \multicolumn{2}{c}{F1-Score} &  \\ \toprule
Category & LR & RF & LR & RF & N \\ \midrule
Distal (Scissors) & \textbf{.527} & .526 & \textbf{.722} & .587 & 380 \\
Writing Tripod (Marker) & \textbf{.595} & .540 & \textbf{.616} & .519 & 380 \\
Medium Wrap (Can) & \textbf{.712} & .631 & \textbf{.697} & .570 & 380 \\
Lateral Tripod (Racquetball) & \textbf{.691} & .614 & \textbf{.771} & .696 & 380 \\ 
\hline
Thumb & .582 & \textbf{.663} & \textbf{.800} & .735 & 48 \\
Index & .510 & \textbf{.519} & \textbf{.327} & .314 & 48 \\
Middle & .500 & \textbf{.631} & .356 & \textbf{.473} & 48 \\
Ring & .511 & \textbf{.544} & \textbf{.196} & .050 & 48 \\
Little & .684 & \textbf{.765} & .625 & \textbf{.702} & 48 \\
\hline
On-Object & .578 & \textbf{.605} & \textbf{.688} & .674 &  402\\
On-Body & \textbf{.625} & .602 & \textbf{.680} & .627 &  402\\
In-Air-Mid & \textbf{.643} & .604 & \textbf{.720} & .660 &  402\\
In-Air-Far & \textbf{.589} & .549 & \textbf{.637} & .575 &  402\\
\hline
Generalist & \textbf{.630} & .598 & \textbf{.711} & .629 & 1520 \\ \bottomrule
\end{tabular}
\end{table}

\begin{table}[]
\caption{Data ablation on grasp-object condition compared to the generalist version of GraspR. The generalist model is trained on data aggregated across all categories.}
\label{tab:ablation}
\begin{tabular}{@{}lccccl@{}}
 & \multicolumn{2}{c}{ROC-AUC} & \multicolumn{2}{c}{F1-Score} &  \\ \toprule
Ablation Category & LR & RF & LR & RF \\ \midrule
Distal (Scissors) & .535 & .529 & \textbf{.718} & \textbf{.655} \\
Writing Tripod (Marker) & .595 & .539 & .668 & .536 \\
Medium Wrap (Can) & \textbf{.641} & \textbf{.627} & .608 & .591 \\
Lateral Tripod (Racquetball) & .579 & .552 & \textbf{.737} & .617 \\ 
\hline
Generalist & .630 & .598 & .711 & .629 \\ \bottomrule
\end{tabular}
\end{table}

\section{Qualitative Analysis}

Participant interviews revealed that factors such as object weight and geometry, frequency of interaction, and user goals changed their preferences. This resonates with prior studies showing that user preference is a complex construct, sensitive to the interaction context~\cite{nielsen1994measuring, tractinsky2000beautiful, toomim2011utility}.

When asked \textit{``What object was the hardest to interact with?''}, six of eight participants mentioned the scissors. Participants explained, \textit{``it was harder to match the orientation (while reaching for the targets)''}---P5, and ``It would be hard to use the scissors and tap (at the same time)''---P6. Surprisingly, P7 mentioned the racquetball as the hardest object to interact with, explaining that ``the grasp depends on the index and thumb. Even a small movement with these fingers loosens the grip.'' This contrast suggests that user goals, such as maintaining grasp stability and utilizing the object's physical affordance while interacting, can shift preferences. P8's choice also diverged from the majority; they stated that the pen was the hardest object because \textit{``the grasp was very different from how I normally hold it,''} which motivates adaptive grasp UIs that accommodate idiosyncratic grasps, even in common objects like a pen. When asked ``What finger was easier to use?'', participants mentioned index or thumb. One participant noted the middle finger was convenient when holding the racquetball. Two participants spontaneously stated their preference for squeezing gestures.

In a hypothetical scenario involving an AR media player interface, participants were asked to interact while holding a cup (represented by a can) in their dominant hand. The interface offered two interaction modalities: a grasp UI using gestures from the 2AFC study and a mid-air tap common in AR interfaces. We asked participants to skip one song and simulate the movements involved in the interaction~\cite{oulasvirta2003understanding}. When asked to skip a song, participants were evenly split between the two modalities. When subsequently asked to skip five songs, two participants who had first chosen the mid-air gesture switched to the grasp UI. However, one participant (P1) maintained their preference for the mid-air gesture but dropped the object to interact, while another (P2) consistently used the mid-air option even when both hands were occupied. In the same simulated scenario, we simulated increases in weight and temperature that favored grasp UI use.

\section{Working Prototype}
\label{sec:walkthrough}

\begin{figure*}
    \centering
    \includegraphics[width=.9\linewidth]{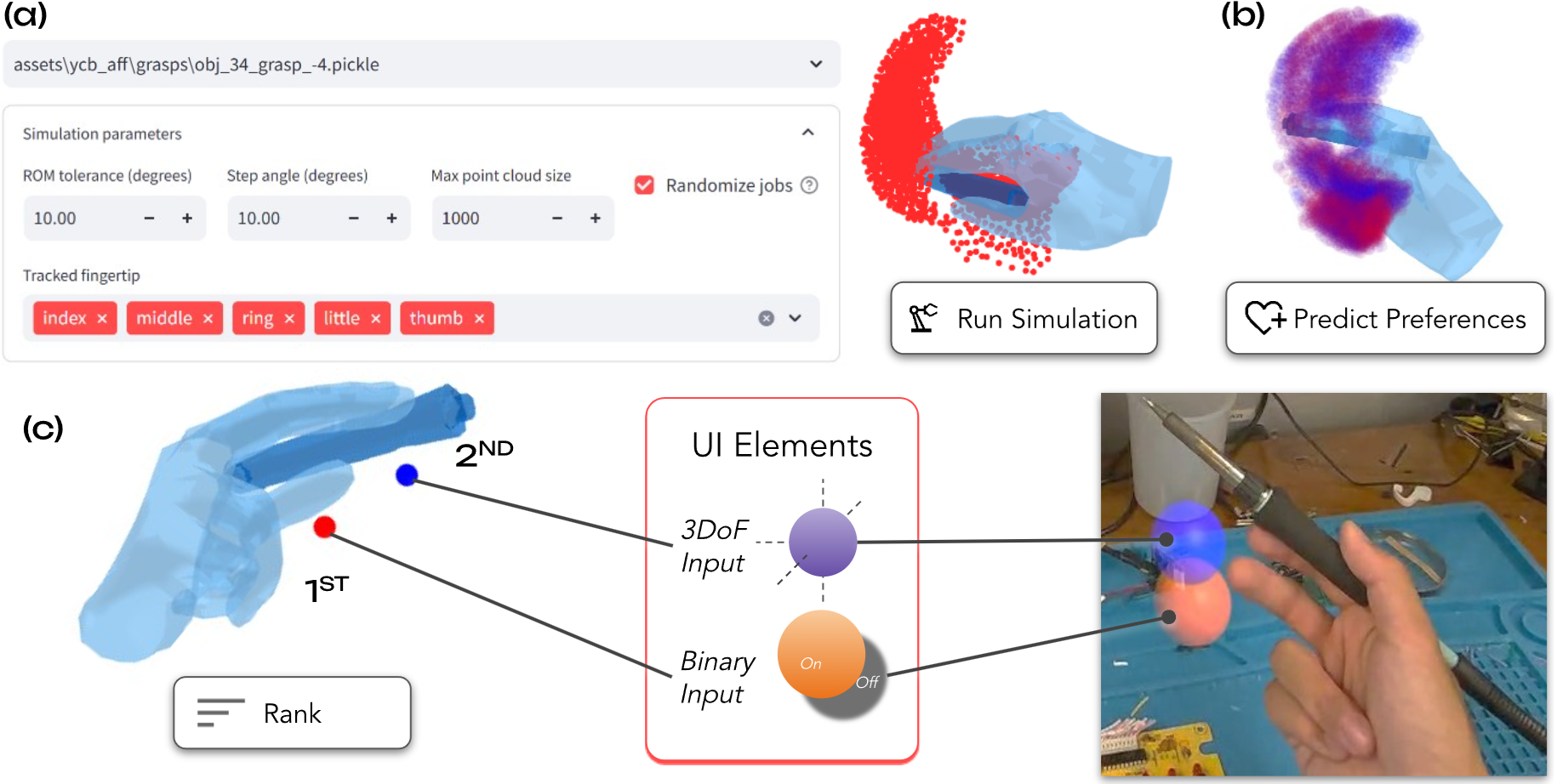}
    \caption{GraspR web UI: (a) select the initial grasp-object condition. You can tune the parameters to control the density and size of the output. Select which fingertips you want to simulate gestures for. Run the simulation to obtain target points within the reach of single-finger grasping gestures. (b) Predict preference for each simulated gesture using GraspR. (c) Rank the gestures by preference and map them to UI elements in your interface with an adaptation schema--i.e., a rule to map single-finger gestures to UI elements. Thanks to the preference predictions and the adaptation schema defined by the designer, the resulting grasp UI can adapt to different grasps and objects.}
    \label{fig:webui}
\end{figure*}

To demonstrate how GraspR enables preference-aware adaptive grasp UIs, we present a design walkthrough~\cite{ledo2018evaluation, tsai2024gait} of the adaptation engine behind the prototypes presented in Section~\ref{sec:prototypes}. Our adaptation process follows three steps: (1) generating a reachable space, (2)  calculating pairwise features, (3) creating a procedural adaptation schema.

GraspR implementation for real-world adaptive grasp interactions requires real-time 3D object reconstruction and MANO grasp representation~\cite{romero2017embodied, yang2021cpf}. Even though real-time hand-object tracking~\cite{wen2023bundlesdf} is an active research field, these capabilities are not yet available in most commercial XR devices. To overcome this technical limitation, we use a Wizard of Oz approach where an observer watches the egocentric video and manually triggers adaptations. 

\paragraph{Forward Kinematic Simulation}
Once the wizard detects the grasp-object condition, we can run the forward kinematic simulation procedure proposed in GraV~\cite{aponte2024grav}. We modified their tool for our purposes by eliminating the dependency on Unity3D. Instead of exporting initial hand and object conditions to Unity for hand armature animation and collision checks, we now directly animate the MANO model~\cite{romero2017embodied, yang2021cpf} provided in the YCB-Affordances dataset~\cite{corona2020ganhand} using a PyTorch implementation~\cite{yang2021cpf} and perform ROM and collision checks with Trimesh~\cite{trimesh}. The results can still be exported as CSV or OBJ files for visualization and development in Unity3D or other development environments. Figure~\ref{fig:webui}a shows the web interface to operate the simulation.

\paragraph{Preference Prediction}
After completing the simulation, we calculate the features described in Section~\ref{sec:feats}. Since GraspR functions as a pairwise classifier, each reachable point and its features must be paired with an alternative. In our walkthrough, we use the initial pose as an anchor, comparing all other poses against it. This approach uses the grasping pose as a reference point to estimate preferences, which can then be applied for adaptations. Figure~\ref{fig:webui}b shows an example of GraspR prediction where stronger preferences are encoded in red.

\paragraph{Adaptation Schema}
Once the pairwise features are calculated, designers can create preference-aware adaptive grasp UIs with GraspR by defining a procedure that associates UI elements with points in the grasp reachable space based on preference scores. In this walkthrough, we consider an interface with two elements: a joystick directional controller and a button. To create the prototypes presented in Section~\ref{sec:prototypes}, our procedure is: (1) to select the point with the highest preference score and associate it with the joystick. (2) To assign the point with the highest preference score from a different finger to the button. This simple procedure creates a preference-aware adaptive layout that not only remains usable through changes in grasp but also maximizes the usage of areas preferred by the user. A simple application can then render the elements as determined by the adaptation procedure, achieving the scenarios presented in Section~\ref{sec:prototypes}. Figure~\ref{fig:webui}c shows a schematic of this adaptation schema followed by the resulting grasp UI.

\section{Discussion, Limitations and Future Work}
\label{sec:discussion}

We now discuss the implications of our results, highlighting limitations and potential for future research on preference modeling for adaptive XR grasp interfaces.

\subsection{Benefits of User Models for XR}

GraspR enables a computational design approach for grasp-based interfaces grounded in empirically measured human preferences. Our model demonstrates three complementary ways that models of user behavior can contribute to XR research and design: \textbf{(1) Enabling adaptive UIs} that respond to contextual changes and maintain alignment with user characteristics. \textbf{(2) Providing efficient proxies} of user behavior to inform rapid exploration of design variations and refine interaction techniques within a computational environment, where a wide range of interaction scenarios can be simulated, reducing the over-reliance on extensive physical prototyping and user studies. \textbf{(3) Distilling actionable guidelines} through explainability techniques allows designers to benefit from the model even before integrating it at runtime. As an active research field~\cite{evangelista2021xrgonomics, zhang2023toward, li2024nicer, li2025alphapig}, user modeling can yield similar benefits through new models and significantly accelerate XR research and product development.

\subsection{Future Perspectives on Data Collection and Modeling}

GraspR establishes the first benchmark in grasp-interaction preference prediction with an ROC-AUC of $.630$ and an F1-Score of $.711$. This demonstrates that spatial user preferences in grasp interactions can be reasonably predicted using simple models with minimal tuning. GraspR performance, while promising, can likely be improved through more sophisticated machine learning techniques and additional data collection.

Expanding the dataset to include more grasp types, specialized tools, and varied object sizes could make the model more robust to preference shifts across different grasp and object conditions. Including underrepresented groups such as children, older adults, and individuals with hand-motor disabilities could enhance our understanding of how grasp preferences vary across user populations. Another promising addition to the dataset would be collecting biomechanical signals such as EMG, which have been used for sensing grasp gestures~\cite{saponas2009enabling, salter2024emg2pose}.

Our study examined short-term, controlled interactions and therefore does not reflect the long-term evolution of user preferences or the impact of real-world contexts. Collecting longitudinal data would enable the development of personalized models that adapt to changes in individual interaction patterns over time. Future research should explore user preferences in ecologically valid settings. Insights from our semi-structured interviews identified initial factors that may influence preferences, including repetition, social context, and user goals. Additional factors may emerge in future research, such as evolving trust, emotional state, cognitive load, life changes, cultural norms, and the influence of social environments. These dynamics highlight the importance of studying preferences over time and across diverse, real-world contexts.

\subsection{Challenges for Real-World Deployment}

GraspR contributes to the broader challenge of dual-task interaction in XR, enabling seamless manipulation of digital content while maintaining engagement with the physical world. As XR technologies integrate into everyday activities, the need for interaction techniques that complement physical tasks becomes paramount.

However, current consumer-grade hardware limitations restrict the precision and responsiveness necessary for practical grasp UI implementation. Current MR headsets rely on vision‑based hand tracking, which often becomes intermittent or fails entirely when the hands are occluded by a handheld object or when lighting is uncontrolled (for example, harsh sunlight or very low indoor illumination). This affects both pose detection and UI stability. Muscle-based sensing in future devices could address this limitation.

Another challenge for real-world deployment stems from the forward kinematic simulation that generates the single-finger reachable points around the grasp. In our study setup, this step required approximately 5 minutes per grasp-object condition, which is orders of magnitude faster than human-based preference elicitation, but still slow enough to reduce responsiveness in real-world deployment. Future work should consider reducing this execution time for the preference prediction of unseen grasp-object conditions. Potential pathways include optimizing the simulation algorithm, approximating results, and predicting preferences directly from object geometry and hand pose.

\section{Conclusion}
We presented GraspR, the first computational model that predicts spatial user preferences for single-finger grasping microgestures, bridging a critical gap between labor-intensive elicitation studies and biomechanical simulations. Our contributions are: (1) the development of GraspR, a model that enables preference-aware adaptive grasp UI design; (2) the release of the first public dataset containing 1,520 user preference comparisons across diverse grasp-object scenarios to support future research; (3) a detailed analysis of design factors, revealing actionable insights—such as favoring thumb and index finger interactions while minimizing wrist flexion; and (4) the demonstration of GraspR’s utility through a working prototype that dynamically adapts interface layouts to user preferences in real-world tasks. GraspR advances the creation of adaptive grasp UIs that prioritize user needs and improve performance in XR contexts. By computationally modeling subjective preferences, our work provides a scalable foundation for designing interfaces that seamlessly integrate physical and digital interactions. 

\begin{acks}
We thank Avinash Nargund for his insightful guidance on rigorous machine-learning evaluation methods. We thank Professor Lixin Yang and Enric Corona for openly sharing the code base and dataset that enabled this work. We thank the U.S. National Science Foundation for supporting this research through the Early CAREER Award 2023 no. 2240133.
\end{acks}

\balance

\bibliographystyle{ACM-Reference-Format}
\bibliography{main}
\end{document}